\documentclass[aps,twocolumn,prl,showpacs,showkeys,showpacs,times,superscriptaddress]{revtex4-1}
\usepackage{amsmath,amssymb,amsthm,times,graphics,graphicx,bm,subfigure}

\usepackage[colorlinks,linkcolor=red,citecolor=blue,urlcolor=blue]{hyperref}
\usepackage{enumerate}
\usepackage{lineno}

\newcommand{\be}{\begin{equation}}
\newcommand{\ee}{\end{equation}}
\newcommand{\ben}{\begin{eqnarray}}
\newcommand{\een}{\end{eqnarray}}
\newcommand{\bes}{\begin{subequations}}
\newcommand{\ees}{\end{subequations}}
\newcommand{\bF}{\begin{figure}}
\newcommand{\eF}{\end{figure}}

\def\pd2v#1#2#3{\frac{\partial^2 #1}{\partial #2 \partial #3}}

\def \2x2mat#1#2#3#4{
\left( \begin{array}{cc}
#1 &  #2 \\  #3 &  #4
\end{array} \right)
}

\begin{document}
\title{Assessing frequency correlation through a distinguishability measurement}
\author{Marco Sbroscia}
\author{Ilaria Gianani}
\author{Emanuele Roccia}
\author{Valeria Cimini}
\author{Luca Mancino}
\author{Paolo Aloe}
\author{Marco Barbieri}
\affiliation{Dipartimento di Scienze, Universit\`a degli Studi Roma Tre, Via della Vasca Navale 84, 00146, Rome, Italy}


\begin{abstract}
The simplicity of a question such as wondering if correlations characterize or not a certain system collides with the experimental difficulty of accessing such information. Here we present a low demanding experimental approach which refers to the use of a metrology scheme to obtain a conservative estimate of the strength of frequency correlations. Our testbed is the widespread case of a photon pair produced per downconversion. The theoretical architecture used to put the correlation degree on a quantitative ground is also described.
\end{abstract}

\maketitle

\section{Introduction}

Time-frequency is the most explored domain in classical communications, however investigations of its quantum counterpart have only recently appeared~\cite{brecht2015}. These have covered a wide range of applications including metrology~\cite{donohue2018,shaofeng2018}, computing~\cite{menicucci2008,pysher2011,yokoyama2013,chen2014}, quantum communications~\cite{nunn2013}. Suitable sources~\cite{ramelow2009,roslund2013,ansari2018} and methods for manipulation~\cite{donohue2013,Manurkar2016,karpinski2016,allgaier2017,ansari2017,davis2017,ra2017,Reddy2018,maclean2018} have been triggered by recent developments. The control of frequency correlations has a central role~\cite{law2000,kuzucu2005,mosley2008,cohen2009,branczyk2011,tischler2015,graffitti2018} and underlies dispersion-free clock synchronization~\cite{giovannetti2001,quan2016} and correlated spectroscopy~\cite{franson2004,Lerch2017} using photon pairs from spontaneous parametric downconversion (SPDC).

A fundamental issue is then to verify and, consequently, to quantify the presence of frequency correlations. A standard procedure makes use of spectrometers at the single photon level: the joint spectral profile of photon pairs is measured with the aim of accessing their correlations. Although feasible in principle, this approach is demanding in terms of resolution particularly when operating SPDC with narrow-bandwidth and, {\it a fortiori}, continuous wave pumps, as correlations manifest on the pump energy scale. This sets hard requirements in practice. 
Such a problem can be partly circumvented by recurring to a stimulated downconversion process~\cite{liscidini2013,eckstein2014,fang2014,rozema2015,Park17}, but the limited gain in the CW condition makes it suitable mostly for four-wave mixing in resonant media, with limited appeal for SPDC. If the sought objective is only to confirm the presence of correlations and put a number on their strength, a less demanding approach would be highly desirable even at the cost of limited detail.

Here we propose a practical method to capture frequency correlations in SPDC photon pairs. It has been demonstrated~\cite{roccia2017a} that phase measurements performed with quantum states are affected by the presence of correlation within the probes. In these conditions, good contrast is obtained even with high dispersion samples. This effect relies on the well-known cancellation of the first-order phase contribution~\cite{franson1992,Baek09}. By this approach it is possible to retrieve a numerical bound for the degree of frequency correlation of the two partner photons in a setup with reduced complexity.

\section{Experiment}

\begin{figure}[b!]
\includegraphics[width=\columnwidth]{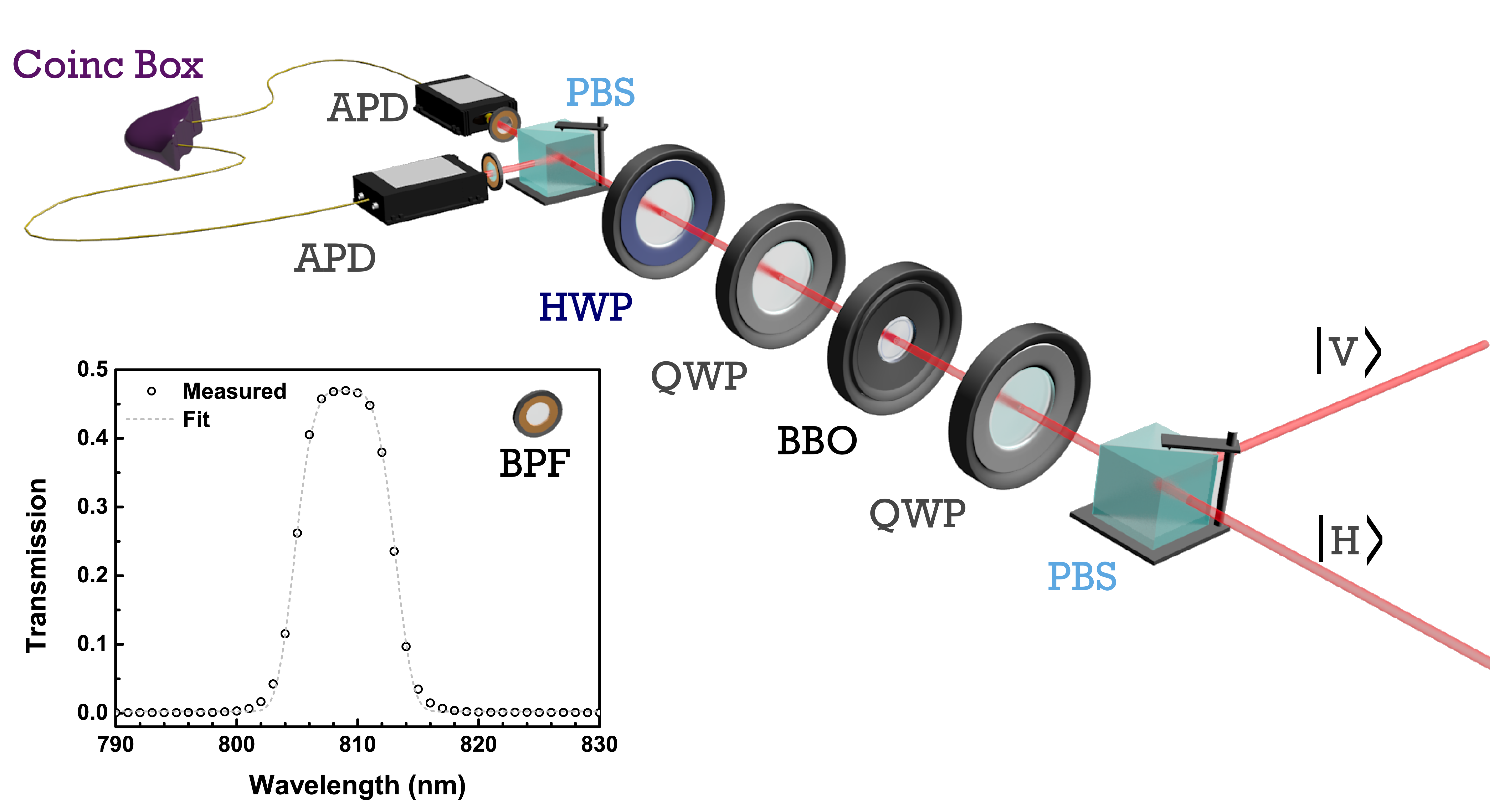}
\caption{{\it Experimental setup: two single photons generated via a type-I SPDC process are driven on the same spatial mode with mutually orthogonal (H/V) polarizations. After letting them pass across a QWP, a 3-mm nominal length BBO crystal and a second QWP, thus imparting an overall phase shift in the circular basis, the emerging quantum state is projected onto linear basis for detection. By collecting coincidence counts between the two output of the setup as a function of the angle $\theta$ of the HWP, oscillations in count rate are recorded. The inset shows the spectral profile of the two bandpass filters (BPF) preceding the photon counters (APD).}}
\label{Fig1}
\end{figure}

Our experiment is sketched in Fig.~\ref{Fig1}: photon pairs are generated from a type-I parametric down-conversion (PDC) source. The process is driven by a cw laser, emitting at central frequency $\Omega_p=405$ nm with a linewidth of the order of 200 MHz, so to deliver photons in the degenerate condition. These two photons, polarised by means of half wave plates (HWPs) in orthogonal directions, are superposed on the same spatial mode on a polarising beam splitter (PBS) with zero delay. When the two photons are produced in indistinguishable separable time-frequency modes, their wavefunction can be written as
\be
|\Psi\rangle=\frac{1}{2}\left({\hat{a}^{\dagger\: 2}_L}-{\hat{a}^{\dagger\: 2} _R}\right)\vert0\rangle,
\label{nun}
\ee
where the operator $\hat{a}^{\dagger}_L$ ($\hat{a}^{\dagger}_R$) describes the creation of a photon in the left-circular (right-circular) polarisation mode: this is then a {\it N}00{\it N} state in the circular polarisation basis.

In order to explore the actual spectral structure of the state and look for the effect of correlations, we introduce a phase delay between the $L$ and $R$ components using a dispersive material. For this purpose we make use of a 3-mm crystal of barium beta borate (BBO), inserted between two quarter wave plates (QWPs) ensuring that the birefringent phase is imparted between the circular components. All optical axes of the plates and the BBO are set at $45^\circ$.  
 
\begin{figure}[t!]
\includegraphics[width=\columnwidth]{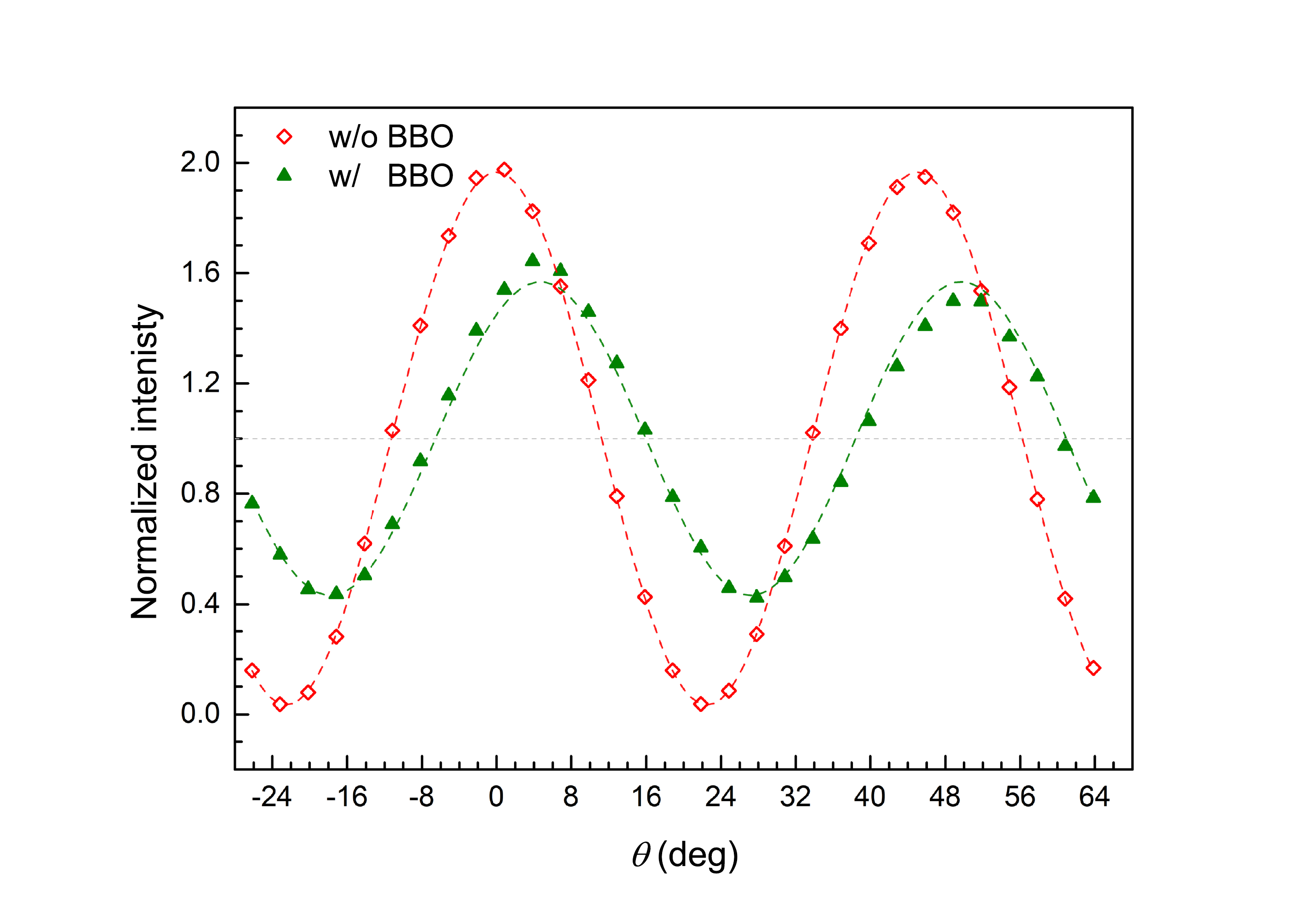}
\caption{{\it Coincidence rate as a function of the angle $\theta$ of the HWP (see Fig~\ref{Fig1}) for the calibration step, without the non linear crystal (red diamonds), and with the crystal inserted (green triangles) to evaluate the imparted phase and the reduction in visibility thus unravelling the correlation level of the photon pair. Intensity has been normalized to have the average value equals to 1. The height of the ridge, or equivalently the depth of the gorge, with respect the mean value measures the visibility of the interferometer, thus the larger this difference the higher the visibility. Dashed lines are the best fit with probability $P(\theta)$ as per Eq.~\eqref{eqn:prob}.}}
\label{Fig2}
\end{figure}

The state is analysed by means of a HWP, set at an angle $\theta$ and a second PBS followed by an avalanche photodiode on each output mode. The frequency detection mode is defined by two interference filters, centred at the degeneracy of the PDC process, with a full width half maximum of 7.3nm (Fig.~\ref{Fig1}, inset). The observed coincidence rate oscillates as $\theta$ is scanned, as shown in Fig.~\ref{Fig2}. 
This reports the behaviours with both the BBO crystal inserted (green solid triangles) and without it (red open diamonds) to have a reference; for the sake of clarity, data have been renormalized scaling their mean value to 1, thus highlighting a phase shift $\phi_0$ in the fringes, and a reduction in their visibility $v$, both due to the presence of the dispersive medium.

Since the interferometer is fed with a two-photon {\it N}00{\it N} state, oscillation frequency is doubled with respect to what expected for a single photon state, thus the phase imparted by the crystal result $\phi_0=0.244$ rad modulus $2\pi$. More interestingly, for what concerns the visibility, it reduces from $v=0.966\pm0.003$ without the crystal to $v=0.568\pm0.012$ once the BBO is in place. Although significant, the reduction in visibility observed with the {\it N}00{\it N} state is much less severe then expected for the case of a single photon probe: given the thickness of the crystal we used, an almost vanishing visibility is expected, as we experimentally checked. This resilience is due to the presence of correlations, to which the measurement of the visibility can give access.

\section{Analysis}

We extend the description of the wavefunction \eqref{nun} to include the frequency domain properly. For this purpose, we introduce a more complete description of the SPDC written as
\begin{equation}
|\Psi\rangle=\int \text{d}\omega_1 \text{d}\omega_2 \Phi (\omega_1, \omega_2) \hat{a}^\dagger _H (\omega_1) \hat{a}^\dagger _V (\omega_2) |0\rangle,
\label{eq:in_state}
\end{equation}
where $\Phi (\omega_1, \omega_2)$, the spectral wavefunction of the pair, is responsible for the correlations. This state evolves under the action of an optically active material presenting dispersion; each photon experiences a frequency-dependent polarisation rotation $\phi(\omega)/2$, due to the different refractive indexes for the $L$ and $R$ polarisations. The analysis is then performed with the HWP at $\theta$. The overall transformation of the polarisation modes is given by:
\be
\begin{aligned}
\hat{a}^\dag_H(\omega)\rightarrow \cos(2\theta+\phi(\omega)/2)\hat{a}^\dag_H(\omega)+\sin(2\theta+\phi(\omega)/2)\hat{a}^\dag_V(\omega)\\
\hat{a}^\dag_V(\omega)\rightarrow \cos(2\theta+\phi(\omega)/2)\hat{a}^\dag_V(\omega)-\sin(2\theta+\phi(\omega)/2)\hat{a}^\dag_H(\omega).
\end{aligned}
\ee
The photons are selected by means of identical frequency filters with spectral shape $f(\omega)$, which can be accurately fit by a 4th-order super-Gaussian curve. Using these formulae, we can write the coincidence detection amplitude at times $t_1$ and $t_2$ as:
\be
A(t_1,t_2)=\tilde\varphi_C(t_1,t_2)-\tilde\varphi_S(t_1,t_2),
\ee
with
\be
\begin{aligned}
\tilde\varphi_C(t_1,t_2)=&\int \text{d}\omega_1 \text{d}\omega_2 e^{i(\omega_1t_1+\omega_2t_2)} \times \\
&\times \Phi (\omega_1, \omega_2)f(\omega_1)f(\omega_2) \cos(\theta_1)\cos(\theta_2),\\
\tilde\varphi_S(t_1,t_2)=&\int \text{d}\omega_1 \text{d}\omega_2 e^{i(\omega_2t_1+\omega_1t_2)} \times \\
&\times \Phi (\omega_1, \omega_2)f(\omega_1)f(\omega_2) \sin(\theta_1)\sin(\theta_2)
\end{aligned}
\ee
and $\theta_i=2\theta+\phi(\omega_i)/2$, $i=1,2$. The actual detection probability is obtained by summing the contributions at all times:
\be
P(\theta)=\int \text{d} t_1 \text{d} t_2 |A(t_1,t_2)|^2
\ee
which contains oscillating terms in $\theta_1-\theta_2$ and $\theta_1+\theta_2$. Notice that the first variable is centred around zero, while the other is centred around $4\theta+\phi(\Omega_0)$, with $\Omega_0=\Omega_p/2$. Its general expression can be evaluated by using Parseval's theorem; this is greatly simplified for the case $\Phi(\omega_1,\omega_2)=\Phi(\omega_2,\omega_1)=\Phi(\omega_1,\omega_2)^*$. These conditions amount to neglect spectral phases due to dispersion and symmetric phase-matching. These deliver the expression
\be 
P(\theta)=\int \text{d} \omega_1 \text{d} \omega_2 |\Phi(\omega_1,\omega_2)|^2 |f(\omega_1)|^2 |f(\omega_2)|^2 \cos(\theta_1+\theta_2)^2,
\label{integrale}
\ee
which is dictated by the correlations between the two photons. 

The dispersion of the medium is taken to be linear in $\omega$, which allows to take the approximation $\theta_1+\theta_2\propto \omega_1+\omega_2$: the probability \ref{integrale} is then more easily calculated in the rotated coordinates $\omega_p=\omega_1+\omega_2$ and $\omega_-=\omega_1-\omega_2$. Furthermore, since the phase-matching bandwidth is typically much larger than the FWHM of the filters, one can neglect the dependence of $\Phi(\omega_1,\omega_2)$ on $\omega_-$ over their support. As for the dependence on $\omega_p$, we take a Gaussian form $\Phi(\omega_p)=2^{-2(\omega_p-\Omega_p)^2/\sigma^2}$, with $\sigma$ its FWHM. When performing the integration $F(\omega_p)=\int \text{d}\omega_- |f(\omega_1)|^2 |f(\omega_2)|^2$, the resulting function $F(\omega_p)$ can be well approximated by a Gaussian with FWHM equal to that of the filters $\delta\omega$ (see Appendix). A measure of the spectral correlation can be then obtained by taking $\sigma^2 = \kappa \delta\omega^2$. Uncorrelated states would have a diverging value of $\kappa$, so that the spectral dependence of the probability $P(\theta)$ is only contained in the filters; in the opposite limit, a perfect correlation would result in $\kappa=0$. The Gaussian shape of the frequency dependence and the linear dispersion approximation, lead to the expression for the probability:
\be
P(\theta)=\frac{1}{2}e^{-\sigma_\phi^2(\kappa)/2}\cos({4\theta+\phi(\Omega_p/2)}),
\label{eqn:prob}
\ee
where $\sigma_\phi^2(\kappa)$ is the variance of the phase distribution, which is expected to take a Gaussian form, and contains information on $\kappa$. The visibility $v$ of the fringes is then governed by the frequency correlation: the experimental value $\bar \kappa$ is thus the one giving $\sigma^2_\phi(\bar \kappa)=-2\log(v)$, estimated in $\bar \kappa=0.14\pm0.02$: this is actually a lower bound on the true value, since we are ascribing to imperfect correlation non-idealities, including those linked to the modelling, as well as experimental artefacts. In this respect, we can observe how the Gaussian shape of the phase distribution is well justified by the low values of skewness ($\mu_3/\sigma_\phi^3 = 0.005$) and kurtosis ($\mu_4/\sigma_\phi^4-3 = 0.014$). The uncertainty on $\bar \kappa$ is mostly dictated by the uncertainty with which the dispersive crystal is placed on the photon path.

\section{Conclusions}
Observing dissipative environments can evidence correlation properties of probe states. We have used this property to assess and quantify the presence of frequency correlations in photon pairs, without recurring to energy-resolved measurements. A numerical bound is obtained under a series of assumptions: the most important is to assume a symmetric wave-function and neglect spectral phases. These are generally well satisfied in CW-pumped SPDC. We should stress that, although genuine frequency entanglement is present, our method is sensitive only to correlations. As a practical comment, no corrections for experimental imperfections have been introduced, {\it e.g.} to take into account the limited contrast observed without the sample. Extensions of our method will need to consider asymmetric and non-degenerate spectra, by means of distinct dispersive media.

\vskip 1mm
{\bf Acknowledgements.} We gratefully thank Marco ``Geno'' Genoni, Fabio Sciarrino, Nicol\'o Spagnolo, Adil Rab and Emanuele Polino for useful discussions. This work has been funded by the European Commission via the Horizon 2020 Programme (Grant Agreement No. 665148 QCUMbER).

\bibliography{Bibliography-2}

\section*{Appendix}
The filter shape in the inset of Fig.~\ref{Fig1} is well described by a super-Gaussian shape of order 4: 
\be 
|f(\omega)|^2=2^{-\left(\frac{2(\omega-\Omega_0)}{\delta \omega}\right)^4}.
\ee 
This is inserted in the general expression of the detection probability \eqref{integrale}, which is then evaluated by using the assumptions on $\Phi(\omega_1,\omega_2)$ introduced in the main text. These lead to neglect its dependence on $\omega_-$, which thus needs to be integrated:
\be
F(\omega_p) =\frac{1}{2} \int \text{d}\omega_- \; \left| f \left( \frac{\omega_p+\omega_-}{2}\right) \right|^2 \left| f \left( \frac{\omega_p-\omega_-}{2} \right) \right|^2.
\ee 
The exact expression of $F(\omega_p)$ is found to be, up to an overall numerical factor:
\be
F(\omega_p) =  |\nu_p| e^{7\nu_p^4} K_{1/4}(9\nu_p^4),
\ee 
with $\nu_p=(\omega_p-\Omega_p)/\delta\omega$, and $ K_{1/4}(x)$ a modified Bessel function of the second kind. 
The distance of the approximating Gaussian distribution to $F(\omega_p)$ can be estimated by the Kullback-Leibler divergence~\cite{kullback1951} $D_{KL}=0.0066$ demonstrating the closeness of the two functions.

\end{document}